\documentclass[manuscript,screen]{acmart}
\renewcommand\footnotetextcopyrightpermission[1]{}

\usepackage{tabularx} 
\usepackage{booktabs}
\usepackage{graphicx} 
\usepackage{booktabs}
\usepackage{multirow}
\usepackage{longtable}
\usepackage{lscape}
\usepackage{array}
\usepackage{ragged2e}
\usepackage{float}
\usepackage{adjustbox}

\setcopyright{none}

\makeatletter
\def\@acmSubmissionID{}
\@twosidefalse
\@mparswitchfalse
\makeatother

\begin{document}

\begin{center}
\textbf{This is a preprint under review at ACM TOIS. Do not redistribute the final version without permission.}
\end{center}
\title[Retrieval-Augmented Generation: A Survey]{Retrieval-Augmented Generation: A Comprehensive Survey of Architectures, Enhancements, and Robustness Frontiers}


\author{Chaitanya Sharma}
\affiliation{%
  \institution{Independent Researcher}
  \country{United States}
}


\begin{abstract}
Retrieval-Augmented Generation (RAG) has emerged as a powerful paradigm to enhance large language models (LLMs) by conditioning generation on external evidence retrieved at inference time. While RAG addresses critical limitations of parametric knowledge storage—such as factual inconsistency and domain inflexibility—it introduces new challenges in retrieval quality, grounding fidelity, pipeline efficiency, and robustness against noisy or adversarial inputs. This survey provides a comprehensive synthesis of recent advances in RAG systems, offering a taxonomy that categorizes architectures into retriever-centric, generator-centric, hybrid, and robustness-oriented designs. We systematically analyze enhancements across retrieval optimization, context filtering, decoding control, and efficiency improvements, supported by comparative performance analyses on short-form and multi-hop question answering tasks. Furthermore, we review state-of-the-art evaluation frameworks and benchmarks, highlighting trends in retrieval-aware evaluation, robustness testing, and federated retrieval settings. Our analysis reveals recurring trade-offs between retrieval precision and generation flexibility, efficiency and faithfulness, and modularity and coordination. We conclude by identifying open challenges and future research directions, including adaptive retrieval architectures, real-time retrieval integration, structured reasoning over multi-hop evidence, and privacy-preserving retrieval mechanisms. This survey aims to consolidate current knowledge in RAG research and serve as a foundation for the next generation of retrieval-augmented language modeling systems.
\end{abstract}


\maketitle

\section{Introduction}

Large Language Models (LLMs) have demonstrated impressive generalization across natural language tasks, but their reliance on static, parametric knowledge remains a fundamental limitation. This restricts their ability to handle queries requiring up-to-date, verifiable, or domain-specific information, often resulting in hallucinations or factual inconsistencies~\cite{gao2023retrieval, lewis2020retrieval}.

Retrieval-Augmented Generation (RAG) addresses this issue by coupling pretrained language models with non-parametric retrieval modules that fetch external evidence during inference. By conditioning generation on retrieved documents, RAG systems offer greater transparency, factual grounding, and adaptability to evolving knowledge bases. These properties have made RAG central to tasks such as open-domain QA, biomedical reasoning, knowledge-grounded dialogue, and long-context summarization.

However, integrating retrieval with generation introduces unique challenges: retrieval noise and redundancy can degrade output quality; misalignment between retrieved evidence and generated text can lead to hallucinations; and pipeline inefficiencies and latency make deployment costly at scale. Moreover, balancing modularity with tight retrieval–generation interaction remains an open architectural trade-off.

In this survey, we first present a high-level taxonomy of RAG architectures based on where core innovations occur—within the retriever, the generator, or through their joint coordination (Section~\ref{sec:taxonomy}). We begin with a background on RAG’s mathematical formulation and components (Section~\ref{sec:formulation}), and then explore advances in retrieval strategies, filtering, and control mechanisms (Section~\ref{sec:enhancements}). We further analyze how RAG systems are benchmarked (Section~\ref{sec:evaluation}), compare prominent frameworks (Section~\ref{sec:comparison}), and conclude with open research challenges and future directions (Section~\ref{sec:future}).

\section{Background and foundations of retrieval-augmented generation}
Retrieval-Augmented Generation (RAG) is a framework that augments large language models (LLMs) with external knowledge access via document retrieval. It builds on the intuition that generating grounded and verifiable responses requires not only parametric knowledge stored in model weights, but also non-parametric access to a dynamic evidence corpus. This section outlines the core components of RAG systems and presents the mathematical formulation that underpins their design.

\begin{figure}[t]
  \centering
  \includegraphics[width=\linewidth]{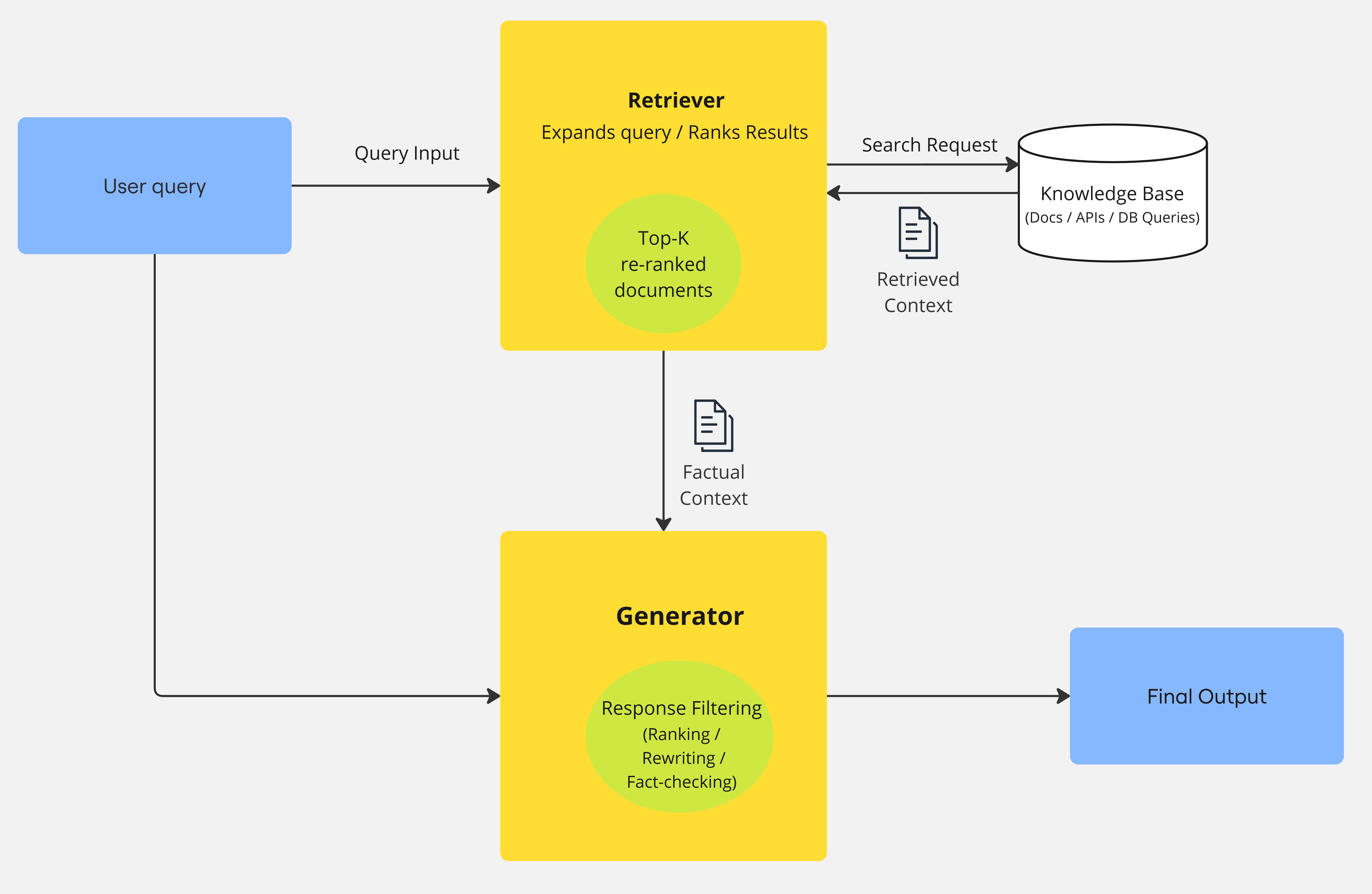}
  \caption{\textbf{Retrieval-Augmented Generation (RAG) workflow.} A user query is processed by the retriever, which may perform query expansion before retrieving documents from external knowledge sources (e.g., databases, APIs, or document stores). Retrieved documents are re-ranked by relevance, and the Top-K are passed to the generator as factual context. The generator synthesizes a response conditioned on both the query and retrieved content. An optional post-processing step (e.g., ranking, rewriting, or fact-checking) may further refine the output, enhancing factual consistency, real-time adaptability, and overall response quality in large language models (LLMs).}
  \label{fig:rag-workflow}
\end{figure}

\subsection{Components of a RAG System}
At a high level, a RAG system consists of three modules:

\textbf{Query Encoder:} Encodes the input $x$ into a query representation $q$, which is used to retrieve relevant documents. This can be either a neural encoder or a rule-based template.

\textbf{Retriever:} Given the query $q$, the retriever fetches a ranked list of documents $d_1, d_2, \ldots, d_k$ from a corpus $\mathcal{C}$. Retrievers may be sparse (e.g., BM25~\cite{robertson2009probabilistic}), dense (e.g., DPR~\cite{karpukhin-etal-2020-dense}), hybrid, or generative.

\textbf{Generator:} The generator conditions on the input $x$ and the retrieved documents $d_i$ to produce the final output $y$. This is typically a pretrained transformer model (e.g., T5~\cite{raffel2020exploring}, BART~\cite{lewis-etal-2020-bart}, GPT~\cite{brown2020language}).

\subsection{Mathematical Formulation}
\label{sec:formulation}
Formally, the generation process in Retrieval-Augmented Generation (RAG) can be expressed as modeling the conditional distribution:

\begin{equation}
P(y \mid x) = \sum_{d \in \mathcal{C}} P(y \mid x, d) \cdot P(d \mid x)
\label{eq:rag-full}
\end{equation}

\noindent where:
\begin{itemize}
  \item $x$ is the input (e.g., a question or prompt),
  \item $d$ is a retrieved document from corpus $\mathcal{C}$,
  \item $y$ is the generated response.
\end{itemize}

In practice, the summation is approximated by retrieving the top-$k$ documents $d_1, \ldots, d_k$, yielding:

\begin{equation}
P(y \mid x) \approx \sum_{i=1}^{k} P(y \mid x, d_i) \cdot P(d_i \mid x)
\label{eq:rag-approx}
\end{equation}

This decomposition reflects two key probabilities:
\begin{itemize}
  \item $P(d_i \mid x)$: the relevance score of document $d_i$ given the input $x$, often derived from a retriever or reranker.
  \item $P(y \mid x, d_i)$: the probability of generating output $y$ conditioned on $x$ and document $d_i$, modeled by the language model.
\end{itemize}

Variants of RAG differ in how they estimate and combine these components. Some use a fixed retriever and let the generator handle noisy inputs, while others jointly optimize retrieval and generation to maximize downstream utility.

\section{Taxonomy of RAG Architectures}
\label{sec:taxonomy}
To contextualize recent advances in Retrieval-Augmented Generation (RAG), we propose a taxonomy that categorizes existing systems based on their architectural focus—retriever-centric, generator-centric, hybrid, and robustness-oriented designs. This classification highlights key design patterns and illustrates how different frameworks tackle the core challenges of retrieval, grounding, and reliability.

\begin{figure}[t]
  \centering
  \includegraphics[width=0.95\linewidth]{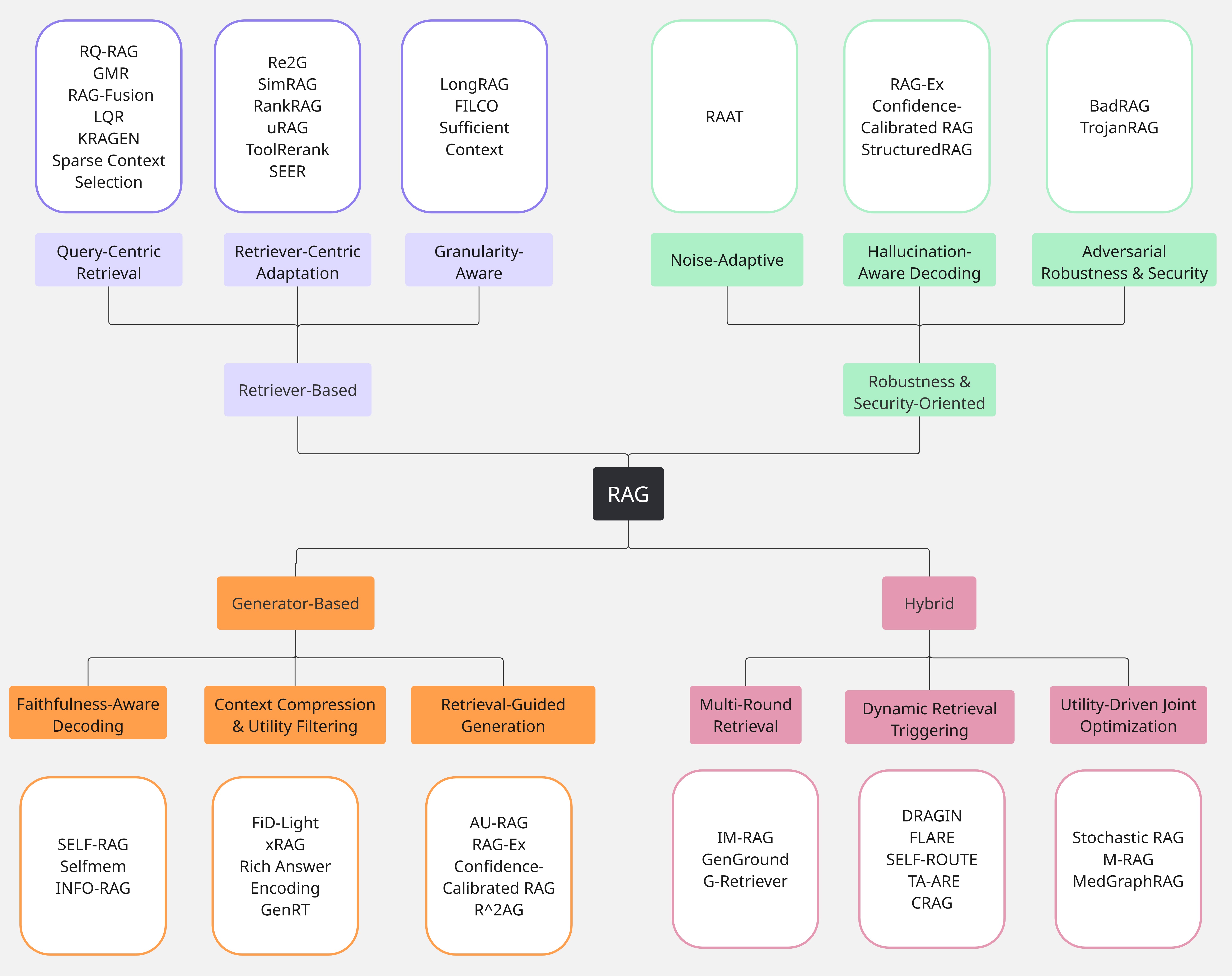}
  \caption{%
    \textbf{Figure 2: Taxonomy of Retrieval-Augmented Generation (RAG) Systems.}
    This taxonomy categorizes RAG frameworks based on their primary architectural orientation—retriever-based, generator-based, hybrid, and robustness-focused designs. Retriever-based models are further grouped into query-centric, retriever-centric, and granularity-aware approaches, while generator-based models include faithfulness-aware decoding, context compression, and retrieval-guided generation. Hybrid systems are organized by retrieval dynamics (e.g., multi-round, utility-driven), and robustness-oriented models address challenges such as noise, hallucination, and adversarial vulnerabilities. This structure highlights the diverse innovations shaping the RAG landscape.
  }
  \label{fig:rag-taxonomy}
\end{figure}

\subsection{Retriever-Based RAG Systems}

Retriever-based Retrieval-Augmented Generation (RAG) systems delegate architectural responsibility primarily to the retriever, treating the generator as a passive decoder. These systems operate under the premise that the fidelity and relevance of the retrieved context are the most critical factors for generating accurate and grounded outputs. Innovations in this space typically fall into one of three design patterns: input-side query enhancement, retriever-side adaptation, and retrieval granularity optimization.

\textbf{Query-Driven Retrieval:} A prominent strategy focuses on refining and structuring user intent before retrieval to maximize alignment with relevant corpus segments. This includes decomposition, rewriting, generative reformulation, and the incorporation of structured priors to guide retrieval. Notable examples include RQ-RAG (Refine Query for RAG)~\cite{chan2024rqrag}, which decomposes multi-hop queries into latent sub-questions, and GMR (Generative Multi-hop Retrieval)~\cite{lee-etal-2022-generative}, which uses a generative LLM to autoregressively formulate complex multi-hop queries. RAG-Fusion~\cite{Rackauckas_2024} further improves recall by combining results from multiple reformulated queries through reciprocal rank fusion~\cite{cormack2009reciprocal}. Structured approaches have also emerged: KRAGEN (Knowledge Retrieval Augmented Generation ENgine)~\cite{matsumoto2024kragen} introduces graph-of-thoughts prompting to decompose complex queries into subproblems, retrieving relevant subgraphs to guide multi-hop reasoning. Additionally, LQR (Layered Query Retrieval)~\cite{app142311014} implements hierarchical planning over multi-hop questions, while Sparse Context Selection~\cite{zhu2025accelerating} emphasizes efficient sparse reformulations for both recall and speed.

\textbf{Retriever-Centric Adaptation:} Another line of work modifies the retriever itself through architectural enhancements or task-specific learning. Re2G (Retrieve, Rerank,
Generate)~\cite{glass-etal-2022-re2g} blends symbolic and neural retrieval via reranking layers, while SimRAG (Self-Improving RAG)~\cite{xu-etal-2025-simrag} employs self-training over synthetic QA pairs to improve domain generalization. Frameworks like RankRAG~\cite{rankrag2024} and uRAG (unified RAG)~\cite{urag2024} emphasize retriever versatility—either by unifying reranking and generation within a single backbone or by training general-purpose retrievers across varied downstream tasks. Additionally, systems such as ToolRerank~\cite{zheng-etal-2024-toolrerank} dynamically adapt retrieval strategies based on query semantics, optimizing tool selection in specialized retrieval settings. Relatedly, SEER (Self-Aligned Evidence Extraction for RAG)~\cite{zhao-etal-2024-seer} focuses on post-retrieval adaptation, advancing corpus-based evidence extraction by aligning evidence selection with faithfulness, helpfulness, and conciseness criteria, thereby improving evidence quality for open-domain and temporally sensitive queries.

\textbf{Granularity-Aware Retrieval:} This pattern addresses retrieval precision by optimizing the unit of retrieval—from full documents to fine-grained, semantically aligned segments. LongRAG~\cite{jiang2024longrag} exemplifies this by retrieving compressed long-context chunks, constructed through document grouping, to better exploit long-context language models. Similarly, FILCO (Filter Context)~\cite{wang2023filco} enhances retrieval granularity by filtering irrelevant or low-utility spans from retrieved passages before generation, improving the faithfulness and efficiency of RAG outputs. In parallel, the Sufficient Context analysis framework~\cite{joren2025sufficient} offers a complementary lens, evaluating whether retrieved contexts contain enough information to support accurate generation, thereby highlighting the importance of granular retrieval quality for system robustness.

Each of these patterns anchors its innovation in the retriever, preserving modularity and interpretability. However, they also introduce trade-offs in latency, redundancy, and sensitivity to ambiguous or underspecified queries. Section~\ref{sec:enhancements} elaborates on how downstream enhancements—such as reranking, adaptive filtering, and utility-based context selection—further address these limitations.

\subsection{Generator-Based RAG Systems}

Generator-based RAG systems concentrate architectural innovation on the decoding process, assuming the retrieved content is sufficiently relevant and shifting the burden of factual grounding and integration to the language model. These systems enhance output quality through mechanisms for self-verification, compression, and controlled generation. We identify three recurring design patterns within this category: faithfulness-aware decoding, context compression and utility filtering, and retrieval-conditioned generation control.

\textbf{Faithfulness-Aware Decoding:} To reduce hallucinations and improve factual grounding, several systems embed mechanisms for self-reflection, verification, or correction during generation. SELF-RAG (Self-Reflective RAG)~\cite{asai2024selfrag} introduces a critique--generate loop, allowing the model to assess and revise its outputs before finalization. SelfMem~\cite{selfmem2023} builds on this by incorporating a self-memory module that enables the model to revisit and refine prior generations. INFO-RAG~\cite{xu-etal-2024-info-rag} treats the LLM as a denoising module trained with contrastive objectives. Collectively, these systems decouple output faithfulness from retrieval fidelity, enabling recovery even when retrieval is suboptimal.

\textbf{Context Compression and Utility Filtering:} To address context window limitations, several systems optimize retrieval inputs into denser or more structured forms. FiD-Light~\cite{fidlight2023}, a streamlined variant of the Fusion-in-Decoder (FiD) architecture~\cite{izacard-grave-2021-fid}, improves decoding efficiency by compressing encoder outputs across retrieved passages and pruning cross-passage attention without modifying retrieval mechanisms. xRAG~\cite{cheng2024xrag} projects document embeddings directly into the model's representation space, minimizing token overhead through modality fusion. Rich Answer Encoding (RAE)~\cite{huang-etal-2023-retrieval} enhances retrieval relevance by embedding answer-aligned semantics into retriever outputs rather than relying on token overlap. GenRT~\cite{xu-2024-genrt} further refines retrieval utility by reranking and dynamically truncating retrieved lists, retaining only the most contextually valuable candidates for generation. Complementing these designs, an information bottleneck-based filtering approach~\cite{zhu-etal-2024-information} selectively preserves evidence most informative for generation. Together, these strategies advance decoding efficiency and context quality, particularly for long-form and multi-hop RAG tasks.

\textbf{Retrieval-Guided Generation:} A third strategy modulates generation based on retrieval metadata, task-specific cues, or agentic decision-making. AU-RAG (Agent-based Universal RAG)~\cite{jang2024aurag} exemplifies this by using an agent to decide dynamically between retrieved and parametric knowledge across diverse data environments. RAG-Ex~\cite{sudhi2024ragex} perturbs retrieval context to analyze how variability influences model behavior and reliance on external evidence. \text{R\textsuperscript{2}AG} (Retrieval information into
RAG)~\cite{ye-etal-2024-r2ag} extends this by recursively reranking candidates during generation, dynamically prioritizing evidence based on the evolving answer state. In high-stakes domains, Confidence-Calibrated RAG~\cite{confcalrag2025} shows that document ordering and prompt structure affect output certainty, highlighting the need for calibration alongside factual accuracy.

These architectures are particularly suited to domains where factual correctness, reasoning transparency, or structured output formats are essential—such as biomedical QA, finance, and enterprise workflows. While they leave the retriever fixed, many of their techniques are complementary to retrieval-side enhancements and can be layered atop other RAG variants. Section~4.2 further explores compression, reranking, and decoding control strategies in these systems.

\subsection{Hybrid RAG Systems}

Hybrid RAG systems tightly couple the retriever and generator, moving beyond modular architectures to treat retrieval and generation as co-adaptive reasoning agents. These systems emphasize iterative feedback, utility-aware coordination, and dynamic control over retrieval actions, particularly in open-domain, multi-hop, and evolving knowledge contexts. We identify three dominant architectural patterns: iterative or multi-round retrieval, utility-driven joint optimization, and retrieval-aware generation control.

\textbf{Iterative or Multi-Round Retrieval:} These systems interleave retrieval and generation across multiple reasoning steps, allowing for evidence refinement and progressive answer construction. IM-RAG (Inner Monologue RAG)~\cite{yang2024imrag} simulates an ``inner monologue'' by alternating between generation and retrieval phases, supporting multi-step reasoning. Generate-Then-Ground (GenGround)~\cite{shi-etal-2024-genground} follows a similar philosophy, generating a provisional answer first and then retrieving supporting evidence to substantiate or revise it—improving factuality and interpretability in multi-hop settings. G-Retriever~\cite{he2024gretriever} retrieves graph-structured subcomponents as generation unfolds, enhancing complex reasoning over textual graphs.

\textbf{Utility-Driven Joint Optimization:} Several frameworks seek to align retriever outputs with their downstream utility for generation through joint objectives or reinforcement learning. Stochastic RAG~\cite{zamani2024stochastic} treats retrieval as an expected utility maximization problem, updating both retriever and generator end-to-end using REINFORCE-based gradients. M-RAG~\cite{wang-etal-2024-mrag} applies multi-agent reinforcement learning, coordinating distributed retrievers and generators via shared memory and task-specific roles. MedGraphRAG~\cite{wu2024medical} integrates knowledge graphs into the joint learning loop, facilitating domain-specific reasoning with structured priors. These systems improve factuality and answer consistency, particularly in biomedical and enterprise domains.

\textbf{Dynamic Retrieval Triggering:} A growing class of systems dynamically controls when and how to retrieve, conditioned on generation uncertainty, task complexity, or intermediate outputs. DRAGIN (Dynamic Retrieval Augmented
Generation based on the Information Needs
of LLMs)~\cite{su-etal-2024-dragin} triggers retrieval at the token level using entropy-based confidence signals, while FLARE (Forward-Looking Active REtrieval augmented
generation ()~\cite{jiang-etal-2023-active} selectively retrieves based on low-confidence predictions during sentence generation. SELF-ROUTE~\cite{li-etal-2024-self-route} dynamically routes tasks between retrieval and generation modules based on model self-assessed difficulty, and AU-RAG~\cite{jang2024aurag} leverages agentic decision-making to mediate between diverse retrieval sources and procedural knowledge. TA-ARE (Time-Aware Adaptive REtrieval)~\cite{zhang-etal-2024-retrievalqa} introduces a retrieval trigger classifier that adaptively determines when retrieval is necessary and adjusts the granularity of evidence based on query needs. A related approach, CRAG (Corrective RAG)~\cite{yan2024corrective}, evaluates retrieved evidence quality before generation and dynamically decides whether to proceed with generation, re-trigger retrieval, or decompose the input into simpler sub-queries. This corrective mechanism positions CRAG within the hybrid class, as it tightly coordinates retrieval assessment with adaptive generation pathways under uncertainty.

These architectures reflect a broader trend toward treating retrieval as a controllable, contextualized act rather than a fixed preprocessing step. Their strength lies in adaptivity, coordination, and the capacity to handle under-specified or evolving queries. However, they introduce new challenges in training stability, latency, and system transparency---especially when retrieval is performed mid-decoding. These trade-offs, as well as efficiency-oriented enhancements like pipelining and reranking, are further explored in Section~4.

\subsection{Robustness and Security-Oriented RAG Systems}

Robustness- and security-oriented RAG systems are designed to preserve output quality in the face of noisy, irrelevant, or adversarially manipulated retrieval contexts. Unlike models that optimize retrieval or generation under ideal assumptions, these systems explicitly address worst-case scenarios---such as hallucination under misleading evidence, retrieval failures, or corpus poisoning. We identify three major design strategies in this category: noise-adaptive training, hallucination-aware decoding constraints, and adversarial robustness.

\textbf{Noise-Adaptive Training Objectives:} These systems aim to make RAG outputs resilient to degraded or spurious input evidence by training under perturbed, irrelevant, or misleading contexts. RAAT~\cite{fang-etal-2024-enhancing} classifies retrieved passages into relevant, irrelevant, or counterfactual categories and introduces an adversarial training objective to maximize worst-case performance. Bottleneck Noise Filtering~\cite{zhu-etal-2024-information} applies information bottleneck theory to identify the intersection of useful and noisy information, compressing retrieved context into minimal, high-utility representations. These approaches are particularly effective in retrieval-heavy pipelines where context precision cannot be guaranteed.

\textbf{Hallucination-Aware Decoding Constraints:} To mitigate factual inaccuracies in generation, several systems introduce decoding-time constraints or architectures designed to enforce grounding. RAGTruth~\cite{niu-etal-2024-ragtruth} provides benchmarks and evaluation protocols for hallucination detection, guiding system-level design. Structured retrieval-based approaches have also been explored: one method~\cite{fidlight2023} retrieves executable templates (e.g., JSON workflows) to constrain output generation, minimizing reliance on generative interpolation and reducing domain-specific hallucination. RAG-Ex~\cite{sudhi2024ragex} simulates retrieval variability by injecting perturbed documents during training, improving robustness to inconsistent or adversarial context. In high-stakes domains such as healthcare, Confidence-Calibrated RAG~\cite{confcalrag2025} explores how document ordering and prompt design affect both answer accuracy and model certainty.

\textbf{Adversarial Robustness and Security:} Emerging work also highlights new vulnerabilities. BadRAG~\cite{xue2024badrag} and TrojanRAG~\cite{cheng2024trojanrag} demonstrate that adversarially poisoned passages can serve as semantic backdoors, triggering specific behaviors in LLM outputs even when base models remain unmodified. These attacks rely on stealthy corpus manipulations that are hard to detect and pose significant threats in open-domain or API-exposed RAG systems.

Collectively, these systems complement retrieval- and generation-oriented architectures by offering essential safety guarantees in real-world deployments. Their robustness strategies—ranging from retrieval verification and context compression to constrained generation—are modular and often integrable into existing RAG pipelines.

\section{Enhancements in RAG}
\label{sec:enhancements}
Recent advancements in Retrieval-Augmented Generation (RAG) increasingly focus on targeted enhancements across the retrieval--generation pipeline. Beyond architectural baselines, these enhancements address key limitations in retrieval quality, context integration, computational efficiency, robustness to perturbations, and ranking precision. This section delineates five core areas of optimization---retrieval, filtering, efficiency, robustness, and reranking---each contributing to the development of more reliable and performant RAG systems, and collectively summarized in Table~\ref{tab:enhancement-summary}, which compares representative methods based on their mechanisms, strengths, limitations, and ideal use cases.

\subsection{Retrieval Enhancement}

RAG systems have increasingly adopted smarter retrieval strategies to mitigate inefficiencies such as redundant lookups, irrelevant context, and computational overhead. These improvements can be categorized into four major families: adaptive retrieval, multi-source retrieval, query refinement, and hybrid or structured retrieval. Each addresses a distinct bottleneck in the retrieval pipeline, offering trade-offs in latency, scalability, and faithfulness.

\textbf{Adaptive retrieval} dynamically adjusts when to retrieve based on model uncertainty or predictive confidence. TA-ARE replaces static thresholds with a learned estimator, reducing redundant retrievals by 14.9\% in short-form tasks. DRAGIN takes this further by applying retrieval at the token level, using entropy signals to detect knowledge gaps and triggering retrieval through a self-attentive query formulation process. Though it improves multi-hop QA precision, DRAGIN introduces notable inference costs, mitigated through adaptive frequency thresholds. FLARE proactively anticipates knowledge needs before uncertainty arises, improving faithfulness but requiring careful thresholding to avoid excessive retrieval.

\textbf{Multi-source retrieval} targets adaptability across evolving corpora or specialized domains. AU-RAG introduces agent-based retrieval, dynamically selecting sources based on metadata heuristics. This improves domain coverage but necessitates hierarchical pipelines to manage source prioritization. SimRAG enhances retrieval precision using self-supervised learning on synthetic QA pairs, filtered via round-trip consistency. While it achieves 1.2--8.6\% accuracy gains across datasets, it risks overfitting, mitigated by human-in-the-loop validation.

\textbf{Query refinement} techniques enhance retrieval relevance by modifying ambiguous or underspecified queries. RQ-RAG uses perplexity-driven decomposition and rewriting to improve relevance, especially in multi-fact scenarios. However, this incurs inference overhead, mitigated through selective refinement based on query ambiguity. \text{R\textsuperscript{2}AG} improves post-retrieval alignment by injecting retrieval metadata into prompts, bridging the retriever--generator semantic gap. Though effective, it adds computational cost, addressed by only enabling metadata prompting when retrieval scores fall below a relevance threshold.

\textbf{Hybrid and structured retrieval} approaches improve coherence by integrating unstructured and structured sources. M-RAG clusters knowledge into semantic partitions, with dual agents selecting and refining content. It reduces noise but introduces latency, mitigated by dynamic partition expansion. KRAGEN retrieves subgraphs from knowledge graphs, using Graph-of-Thoughts prompting for relational reasoning. This reduces hallucinations by 20--30\%, though it increases memory overhead, controlled via selective node expansion.

Extending hybrid retrieval designs, the Dual-Pathway KG-RAG framework~\cite{xu2024kgrag} combines structured retrieval from knowledge graphs with unstructured corpus retrieval in parallel, enhancing factual consistency and reducing hallucinations by 18\% in biomedical QA tasks. Similarly, Graph RAG~\cite{edge2025localglobalgraphrag} constructs entity-centric graphs from retrieved passages and uses community summarization to scale RAG to large corpora, improving multi-hop QA recall by 6.4 points compared to baseline retrieval. Likewise, Customer Service QA~\cite{xu2024customer} integrates RAG with knowledge graphs constructed from issue-tracking tickets, achieving a 77.6\% improvement in retrieval MRR and a 28.6\% reduction in resolution time when deployed at LinkedIn’s customer service team.

In a complementary direction, Doan et al.~\cite{doan-etal-2024-hybrid} propose a lightweight hybrid retrieval strategy that combines unstructured text embeddings with structured knowledge graph embeddings without requiring complex retriever re-training, achieving up to 13.1\% improvements in retrieval correctness and ranking precision in domain-specific RAG deployments.

\subsection{Enhancing Context Relevance through Filtering}

Despite advances in retrieval models, RAG systems often integrate irrelevant, redundant, or semantically noisy documents that degrade generation quality. Filtering techniques aim to reduce hallucinations and improve answer relevance by selecting only contextually appropriate content. These methods vary in supervision, granularity, and efficiency, and can be categorized into three groups: lexical/statistical filters, information-theoretic optimizers, and self-supervised passage scoring.

\textbf{Lexical filters} such as FILCO apply word overlap and statistical relevance scoring. Using STRINC and CXMI metrics, FILCO removes low-relevance passages and reduces hallucinations by up to 64\%, improving EM by +8.6. However, its reliance on lexical similarity limits its adaptability across domains and query styles.

\textbf{Information-theoretic methods} like IB Filtering~\cite{zhu-etal-2024-information} use principles from the information bottleneck framework to retain only high-utility input features while discarding noise. Though computation-heavy, IB Filtering improves EM by +3.2 with a 2.5\% compression ratio, offering a balance between precision and conciseness. Similarly, Stochastic Filtering models retrieval as an expected utility maximization problem and re-ranks passages based on marginal value, achieving consistent retrieval effectiveness gains with minimal retriever changes.

\textbf{Self-supervised methods} like SEER and RAG-Ex use internal feedback signals to filter noisy retrievals. SEER applies label-free training and generates pseudo-relevance judgments, improving F1 by 13.5\% and achieving a 9.25× reduction in context length. RAG-Ex perturbs retrieved passages and compares generation outcomes, selecting those that maximize semantic consistency. It aligns with human-assessed faithfulness 76.9\% of the time and is model-agnostic, though computationally expensive due to multiple inference passes.

Collectively, these methods balance retrieval compression, answer faithfulness, and domain adaptability. While lexical filters are efficient, self-supervised models provide deeper semantic filtering and support long-form reasoning.

\subsection{Efficiency Enhancements}

While Retrieval-Augmented Generation (RAG) significantly enhances factual consistency in large language models (LLMs) by integrating external document retrieval, it introduces new inefficiencies. These include increased memory overhead, latency from retrieval-processing pipelines, and redundancy in passage selection. This section synthesizes key research efforts aimed at improving retrieval efficiency across four areas: sparse retrieval and context selection, inference acceleration, caching and redundancy reduction, and retrieval faithfulness.

\textbf{Sparse context selection and retrieval-aware generation} techniques aim to reduce the input length and improve semantic alignment without sacrificing output quality. Sparse RAG addresses this by filtering low-relevance content before self-attention, retaining only high-signal tokens via parallel encoding. While it builds on Fusion-in-Decoder (FiD), it improves efficiency by avoiding dense input concatenation. However, it may discard useful context under suboptimal retrieval, requiring fine-tuning to maintain robustness. \text{R\textsuperscript{2}AG} takes a complementary approach by embedding retrieval representations directly into the LLM's context space, enhancing semantic alignment. Unlike prompt-based methods (e.g., REPLUG~\cite{shi-etal-2024-replug}), \text{R\textsuperscript{2}AG} bypasses explicit concatenation, reducing redundant processing. Both approaches enhance efficiency at different stages but require retriever fine-tuning and increase model complexity.

\textbf{Inference acceleration} strategies focus on reducing decoding latency in autoregressive models by minimizing redundant token processing. FiD-Light achieves this through token-level passage compression, which lowers decoding time while preserving key information. Though effective, aggressive filtering can marginally reduce retrieval precision. Speculative Pipelining~\cite{wang2025speculative} further reduces latency by overlapping retrieval and generation. It incrementally processes top-$k$ candidates before retrieval completes, lowering time-to-first-token (TTFT) by 20--30\%. However, it risks speculative hallucinations unless controlled by fallback mechanisms and selective decoding checkpoints. This line of work opens the door for future speculative decoding architectures---discussed in Section~8---that balance responsiveness and reliability in low-latency applications.

\textbf{Caching and redundancy reduction} techniques aim to eliminate recomputation overhead in repetitive or high-throughput workloads. RAGCache~\cite{jin2024ragcache} introduces a hierarchical caching system that stores key-value tensors from prior retrievals. PGDSF extends this with prefix-aware eviction that prioritizes frequent and important documents. While these methods significantly improve efficiency in common-query settings, their impact diminishes on long-tail distributions and introduces cache complexity.

\textbf{Retrieval faithfulness and answer relevance} methods go beyond lexical similarity to ensure that retrieved documents are factually aligned with the generated output. Rich Answer Encoding (RAE) addresses this using a Retriever-as-Answer Classifier (RAC) and Dense Knowledge Similarity (DKS), which rescore documents based on their plausibility. RAE reduces hallucinations and improves grounding but requires retriever retraining, increasing cost.

Taken together, these optimization strategies enhance efficiency across the RAG pipeline: Sparse RAG and \text{R\textsuperscript{2}AG} improve alignment between retrieved documents and generation; FiD-Light and Speculative Pipelining reduce latency during inference; RAGCache and PGDSF minimize recomputation in high-throughput environments; and RAE advances retrieval faithfulness. Collectively, they represent a move toward more scalable, accurate, and computationally efficient RAG systems.

\subsection{Enhancing Robustness}
\label{subsec:robustness}
RAG systems improve factual accuracy in language models by retrieving external information. However, they remain vulnerable to retrieval noise, hallucinations, and adversarial attacks. While past research has addressed these challenges separately---such as noise resilience, hallucination control, and retrieval security---a unified perspective is essential. This section groups robustness techniques into three areas: noise mitigation, hallucination reduction, and security defenses.

Empirical studies further support this need for a unified view; a recent study identifies seven recurrent failure points in operational RAG systems, spanning retrieval errors, context consolidation failures, hallucinated outputs, and incomplete answers~\cite{sevenfailures2024}.

\textbf{Noise mitigation} strategies target irrelevant, misleading, or adversarial content that can degrade RAG accuracy. However, recent work challenges the assumption that all retrieval noise is detrimental; Cuconasu et al.~\cite{powerofnoise2024} demonstrate that carefully positioned random documents can paradoxically improve LLM reasoning and answer quality by promoting evidence selection behaviors. Two contrasting approaches address this: RAAT and CRAG. RAAT uses adversarial pretraining to expose models to subtle and counterfactual retrieval noise, improving F1/EM scores by 20--30\%. Its high training cost limits it to static, high-stakes domains. CRAG filters low-confidence retrievals at inference time and works well in real-time systems, reducing retrieval errors by 12--18\%. However, it struggles with ``soft noise,'' where superficially relevant content misleads the model.

\textbf{Hallucination reduction} techniques such as Structured RAG~\cite{ayala-bechard-2024-reducing} and IM-RAG aim to improve the faithfulness of generated content. Structured RAG constrains retrieval to verified corpora, lowering hallucination rates by 30--40\% with minimal compute cost. Its drawback is poor adaptability, requiring manual updates. IM-RAG uses iterative retrieval refinement, achieving +5.3 F1 / +7.2 EM on HotPotQA. Though more accurate in evolving domains, it is computationally intensive and slower at inference.

\textbf{Security defenses} focus on adversarial threats such as data poisoning and backdoor attacks. Research into BadRAG shows that poisoning just 0.04\% of a corpus can lead to a 98.2\% attack success rate and 74.6\% system failure. Defenses like cryptographic document signing or adversarial filtering are only partially effective. TrojanRAG embeds backdoors in retrieval embeddings, bypassing traditional sanitization. Stronger mitigations---secure training and integrity validation---are needed but require proactive design. Beyond adversarial attacks, privacy vulnerabilities in RAG systems have also been identified; Zeng et al.~\cite{zeng-etal-2024-good} show that both retrieval databases and pretraining corpora can be exploited through structured prompting, although retrieval can paradoxically help reduce memorization leakage by acting as a grounding mechanism.

\subsection{Enhancements and Optimizations in Reranking}

Reranking plays a vital role in improving the relevance and faithfulness of Retrieval-Augmented Generation (RAG) outputs. While initial retrieval stages often return noisy results, reranking refines document ordering before passage selection and generation, reducing hallucinations and improving response accuracy. Recent work advances reranking across three key areas: adaptive reranking, unified pipelines, and fusion-based reranking.

\textbf{Adaptive reranking} methods dynamically adjust the number of documents reranked based on query complexity. RLT~\cite{meng2024rlt} uses ranked list truncation to improve MRR/nDCG while reducing retrieval noise by 15\%. ToolRerank further adapts reranking depth based on familiarity with seen vs. unseen tools, boosting recall by 12\% in hierarchical retrieval tasks. These methods optimize computation by avoiding unnecessary reranking in low-complexity scenarios.

\textbf{Unified reranking pipelines} combine retrieval, document ranking, and generation within a single architecture. RankRAG fine-tunes a language model to jointly score documents and generate answers, improving MRR@10 by 7.8\% while reducing latency. uRAG extends this to multiple tasks---like QA and fact verification---using shared reranking logic and user-feedback signals, improving cross-task generalization by 8\% MRR@10. These approaches eliminate the overhead of separate ranking modules and increase retrieval consistency.

\textbf{Fusion-based reranking} strategies aggregate evidence from multiple query variants to improve answer robustness. RAG-Fusion generates multiple subqueries and applies reciprocal rank fusion, improving answer accuracy by 9\%. \text{R\textsuperscript{2}AG} refines these rankings iteratively, reducing irrelevant retrievals by 15\% through recursive feedback. These models are especially effective for multi-hop and ambiguous tasks.

Reranking methods significantly boost the efficiency and faithfulness of RAG systems. Future work may explore hybrid approaches combining adaptive truncation with fusion-based aggregation, as well as domain-adaptive reranking for enterprise scalability. As RAG expands to more tasks and domains, reranking will remain essential to enabling context-aware, trustworthy generation.

\begin{table*}[t]
\centering
\caption{\textbf{Summary of RAG System Enhancements.} This table categorizes enhancements across five dimensions—retrieval, filtering, efficiency, robustness, and reranking. Each entry specifies the enhancement type, method, mechanism, key strengths, known limitations, and ideal use cases.}
\label{tab:enhancement-summary}
\small
\renewcommand{\arraystretch}{1.2} 
\resizebox{\textwidth}{!}{%
\begin{tabular}{p{1.5cm} p{2cm} p{1.5cm} p{3.5cm} p{3.5cm} p{3.5cm} p{3.5cm}}
\toprule
\textbf{Enhancement Type} & \textbf{Category} & \textbf{Method} & \textbf{Mechanism} & \textbf{Strengths} & \textbf{Limitations} & \textbf{Best Use Case} \\
\midrule
\multirow{15}{*}{Retrieval} & Adaptive & TA-ARE & Dynamic confidence estimation & Reduces redundant retrieval & Estimator latency & Short-form QA \\
& Adaptive & DRAGIN & Token-level entropy-based triggers & Improves multi-hop QA precision & High inference cost & Multi-hop QA \\
& Adaptive & FLARE & Preemptive uncertainty detection & Enhances faithfulness & Risk of over-retrieval & Long-form generation \\
& Multi-source & AU-RAG & Agent-based source selection & High domain adaptability & Source management overhead & Evolving corpora \\
& Multi-source & SimRAG & Synthetic QA + round-trip filtering & Cross-domain accuracy gains & Overfitting risk & Specialized domains \\
& Query & RQ-RAG & Perplexity-based query rewriting & Improves query clarity and relevance & Additional inference steps & Multi-fact queries \\
& Query & R\textsuperscript{2}AG & Retrieval-aware prompt injection & Enhances factual grounding & Prompt expansion overhead & Low-confidence queries \\
& Hybrid & M-RAG & Semantic partitioning + dual agents & Reduces retrieval noise & Partition latency & Context-heavy reasoning \\
& Hybrid & KRAGEN & Knowledge graph subgraph retrieval & Improves structured reasoning & Memory and compute intensive & Biomedical, graph-based tasks \\
\midrule
\multirow{9}{*}{Filtering} & Lexical & FILCO & STRINC + CXMI scoring & +8.6 EM, 64\% hallucination reduction & Query-style bias & Structured QA \\
& Info-Theoretic & IB Filtering & Bottleneck-based compression & +3.2 EM, 2.5\% compression & Computation overhead & High-precision QA \\
& Info-Theoretic & Stochastic Filtering & Utility-maximizing re-ranking & Improves effectiveness & Needs custom scoring & Lightweight retrieval tasks \\
& Self-Supervised & SEER & Pseudo-relevance via self-training & +13.5\% F1, 9.25x context reduction & High training cost & Open-domain QA \\
& Self-Supervised & RAG-Ex & Generation perturbation comparison & 76.9\% human-aligned faithfulness & Multiple inference passes & Faithful generation \\
\midrule
\multirow{11}{*}{Efficiency} & Sparse Selection & Sparse RAG & Retains high-signal tokens & Reduces memory, improves relevance & May discard useful docs & Long-context tasks \\
& Sparse Selection & R\textsuperscript{2}AG & Context-aware retrieval injection & Enhances coherence, lowers redundancy & Retriever fine-tuning needed & Knowledge-intensive QA \\
& Inference Acceleration & FiD-Light & Compresses passages & Faster decoding & Slight loss in recall & Low-latency applications \\
& Caching & Speculative Pipelining & Overlaps retrieval and generation & 20–50\% TTFT reduction & Risk of hallucination & Real-time applications \\
& Caching & RAGCache & Hierarchical cache w/ PGDSF & Eliminates recomputation & Cache complexity in long-tail & High-throughput workloads \\
& Retrieval Quality & RAE & Retriever-as-answer scorer & Boosts grounding and precision & Requires scoring/retraining & Factual QA \\
\midrule
\multirow{9}{*}{Robustness} & Noise Mitigation & RAAT & Adversarial training & +20--30\% F1/EM & High training cost & Offline pretraining \\
& Noise Mitigation & CRAG & Inference-time filtering & +12--18\% precision gain & Ineffective on “soft” noise & Real-time support \\
& Hallucination Control & Structured RAG & Curated corpus retrieval & 30–40\% hallucination reduction & Low adaptability & Static domains \\
& Hallucination Control & IM-RAG & Iterative retrieval refinement & +5.3 F1 / +7.2 EM & Inference latency & Multi-hop QA \\
& Security & BadRAG & Adversarial retrieval poisoning & Demonstrates corpus-level threat & Needs stronger filtering & Security evaluation \\
& Security & TrojanRAG & Embedding-level backdoor & Persistent attack vector & Requires secure training & Security-sensitive pipelines \\
\midrule
\multirow{9}{*}{Reranking} & Adaptive & RLT & Dynamic list truncation & +15\% noise reduction & Heuristic tuning needed & Real-time QA \\
& Adaptive & ToolRerank & Familiarity-aware reranking & +12\% recall for unseen tools & Complexity for unseen/frequent tools & Tool-aware retrieval \\
& Unified Pipeline & RankRAG & Joint rerank + generate & +7.8\% MRR@10 & Domain-specific tuning & End-to-end QA systems \\
& Unified Pipeline & uRAG & Shared reranking engine & +8\% MRR@10, task generalization & Higher setup cost & Multi-task enterprise RAG \\
& Fusion-based & RAG-Fusion & Reciprocal rank fusion & +9\% accuracy & Query explosion risk & Complex multi-hop QA \\
& Fusion-based & R\textsuperscript{2}AG & Recursive reranking refinement & 15\% irrelevant retrieval reduction & Higher latency & Iterative reasoning \\
\bottomrule
\end{tabular}
}
\end{table*}

\section{Comparative Analysis}
\label{sec:comparison}
To assess the empirical effectiveness of design innovations in Retrieval-Augmented Generation (RAG), this section presents a comparative analysis of representative frameworks across three key evaluation settings: short-form question answering, multi-hop reasoning, and robustness under retrieval perturbations. Results are reported as relative improvements over both raw and retrieval-augmented baselines, normalized for model and dataset variability. Additionally, we review ablation studies from the literature to disentangle the contributions of specific components such as retrieval triggers, filtering layers, reranking mechanisms, and robustness modules. These insights offer a clearer understanding of which enhancements most significantly impact performance, faithfulness, and efficiency across diverse RAG configurations.

\begin{table*}[t]
\centering
\caption{\textbf{Comparative Performance of Retrieval-Augmented Generation Frameworks Across Multi-Hop and Short-Form QA Benchmarks.} This table reports relative performance improvements achieved by each RAG framework over two baselines: (i) the raw backbone language model (B) and (ii) the same model augmented with a standard retrieval module (B+R). Results are shown across multi-hop benchmarks (HotpotQA~\cite{yang-etal-2018-hotpotqa}, 2Wiki~\cite{ho-etal-2020-constructing}, MuSiQue~\cite{trivedi-etal-2022-musique}) and short-form QA datasets (PopQA~\cite{mallen-etal-2023-trust}, TriviaQA~\cite{joshi-etal-2017-triviaqa}, ARC-Challenge~\cite{clark2018arc}, NQ~\cite{kwiatkowski-etal-2019-natural}), with metrics including F1, Exact Match (EM), and Accuracy (Acc). Frameworks are grouped by architectural category: retriever-based, generator-based, and hybrid. A ``--'' indicates that the corresponding score was not reported in the original publication. Backbone LLMs referenced in this table include LLaMA 2~\cite{touvron2023llama2}, LLaMA 3~\cite{grattafiori2024llama3}, GPT-3.5/4~\cite{openai2024gpt4}, Vicuna~\cite{vicuna2023}, Mistral~\cite{jiang2023mistral7b}, Mixtral~\cite{jiang2024mixtralexperts}, Gemini~\cite{google2023gemini}, and Gemma~\cite{gemmateam2024}.}
\resizebox{\textwidth}{!}{%
\begin{tabular}{llccccccc}
\toprule
\textbf{Framework} & \textbf{Backbone} & \textbf{HotpotQA} & \textbf{2Wiki} & \textbf{MusiQue} & \textbf{PopQA} & \textbf{TriviaQA} & \textbf{ARC-Challenge} & \textbf{NQ} \\
\midrule
 & & \textbf{B/B+R} & \textbf{B/B+R} & \textbf{B/B+R} & \textbf{B/B+R} & \textbf{B/B+R} & \textbf{B/B+R} & \textbf{B/B+R} \\
\multicolumn{9}{l}{
\textit{Retriever-Based RAG}} \\
RQ-RAG & LLaMA2-7B & 8.485/2.749 (F1) & 1.8/1.396 (F1) & 12.9/4.635 (F1) & 2.884/0.434 (Acc) & -/- & 2.133/1.379 (Acc) & -/- \\
SimRAG & LLaMA3-8B & -/- & -/- & -/- & -/- & -/- & 0.145/- (Acc) & -/- \\
SimRAG & Gemma2-27B & -/- & -/- & -/- & -/- & -/- & 0.034/- (Acc) & -/- \\
SEER & LLaMA2-7B-Chat & 0.104/0.037 (F1) & -/- & -/- & -/- & -/- & -/- & -/- \\
RankRAG & LLaMA3-8B & -/0.079 (F1) & -/0.323 (F1) & -/- & -/- & -/- & -/- & -/- \\
RankRAG & LLaMA3-70B & -/0.242 (F1) & -/0.376 (F1) & -/- & -/- & -/- & -/- & -/- \\
LQR & LLaMA3-8B & 2.081/0.516 (F1) & 0.706/0.141 (F1) & 2.922/0.841 (F1) & -/- & -/- & -/- & -/- \\
LongRAG & GPT-4o & 0.517/- (EM) & -/- & -/- & -/- & -/- & -/- & -/- \\
LongRAG & Gemini-1.5-Pro & 0.696/- (EM) & -/- & -/- & -/- & -/- & -/- & -/- \\
FILCO & LLaMA2-7B & -/0.057 (EM) & -/- & -/- & -/- & -/0.056 (EM) & -/- & -/0.298 (EM) \\
Re2G & BART Large & -/- & -/- & -/- & -/- & 0.251/- (Acc) & -/- & 0.144/- \\

\midrule
\multicolumn{9}{l}{\textit{Generator-Based RAG}} \\
xRAG & Mistral-7B & 0.26/-0.122 (EM) & -/- & -/- & -/- & 0.152/-0.002 (EM) & -/- & 0.293/-0.085 (EM) \\
xRAG & Mixtral-8x7B & 0.207/-0.087 (EM) & -/- & -/- & -/- & 0.043/0.054 (EM) & -/- & 0.126/0.047 (EM) \\
INFO-RAG & LLaMA2-7B & 0.182/- (EM) & -/- & 0.163/- (EM) & -/- & -/- & -/- & -/- \\
INFO-RAG & LLaMA2-13B & 0.222/- (EM) & -/- & 0.358/- (EM) & -/- & -/- & -/- & -/- \\
INFO-RAG & LLaMA2-13B-chat & 0.011/- (EM) & -/- & 0.018/- (EM) & -/- & -/- & -/- & -/- \\
SELF-RAG & LLaMA2-7B & -/- & -/- & -/- & 2.735/0.437 (Acc) & 1.177/0.562 (Acc) & 2.092/0.404 (Acc) & -0.505/- (Acc) \\
SELF-RAG & LLaMA2-13B & -/- & -/- & -/- & 2.796/0.221 (Acc) & 0.8/0.474 (Acc) & -/- & -/- \\
FiD-Light & FiD+DPR & -/- & -/- & -/- & -/- & 0.185/- (EM) & -/- & 0.27/- (EM) \\
R$^2$AG & LLaMA2-7B & 3.231/- (F1) & 34.52/4.445 (F1) & -/- & -/- & -/- & -/- & 0.824/- (Acc) \\

\midrule
\multicolumn{9}{l}{\textit{Hybrid RAG}} \\
DRAGIN & LLaMA2-7B-chat & 0.218/0.338 (F1) & 0.311/0.148 (F1) & -/- & -/- & -/- & -/- & -/- \\
DRAGIN & LLaMA2-13B-chat & 0.368/0.144 (F1) & 0.445/0.169 (F1) & -/- & -/- & -/- & -/- & -/- \\
DRAGIN & Vicuna-13B-v1.5 & 0.279/0.179 (F1) & 0.575/0.371 (F1) & -/- & -/- & -/- & -/- & -/- \\
FLAREdirect & GPT-3.5 & -/- & 0.622/0.223 (F1) & -/- & -/- & -/- & -/- & -/- \\
FLAREinstruct & GPT-3.5 & -/- & 0.353/0.02 (F1) & -/- & -/- & -/- & -/- & -/- \\
GenGround & GPT-3.5 & 0.236/0.093 (F1) & 0.219/0.122 (F1) & 0.359/0.361 (F1) & -/- & -/- & -/- & -/- \\
Stochastic RAG & FiD-Light (T5-Base) & 0.066/- (F1) & -/- & -/- & -/- & -/0.036 (EM) & -/- & -/0.013 (EM) \\
Stochastic RAG & FiD-Light (T5-XL) & 0.065/- (F1) & -/- & -/- & -/- & -/0.016 (EM) & -/- & -/0.037 (EM) \\
CRAG & LLaMA2-7B & -/- & -/- & -/- & 3.034/0.471 (Acc) & -/- & 1.514/0.173 (Acc) & 0.045/- (Acc) \\
Self-CRAG & LLaMA2-7B & -/- & -/- & -/- & 3.204/0.533 (Acc) & -/- & 2.083/0.439 (Acc) & -/- \\
TA-ARE & GPT-3.5 & -/- & -/- & -/- & -/- & -/- & -/- & -/- \\
TA-ARE & GPT-4 & -/- & -/- & -/- & -/- & -/- & -/- & -/- \\
TA-ARE & LLaMA2-7B & -/- & -/- & -/- & -/- & -/- & -/- & -/- \\
\bottomrule
\end{tabular}
}
\label{tab:comparativeanalysis}
\end{table*}

\subsection{Comparative Analysis of Framework Performance on Short-Form QA}

This section presents a comparative analysis of Retrieval-Augmented Generation (RAG) frameworks in short-form question answering, emphasizing their relative improvements over raw large language model (LLM) baselines and retrieval-augmented baselines. As shown in Table~\ref{tab:comparativeanalysis}, these comparisons focus on relative gains (e.g., a value of 2.7 indicates a 270\% improvement) rather than absolute performance metrics, which normalize for variations in backbone architectures, prompting strategies, and evaluation protocols. This approach enables a meaningful comparison across diverse experimental setups.

Among generator-based RAG systems primarily optimized for accuracy, SELF-RAG consistently demonstrates substantial gains across multiple datasets. It achieves over a 270\% improvement from the raw LLM baseline on PopQA~\cite{mallen-etal-2023-trust} and over 200\% on ARC-Challenge~\cite{clark2018arc}, illustrating the effectiveness of deep context integration for enhancing short-form factual recall. FiD-Light, although also a generator-side enhancement, adopts a different optimization philosophy centered on lightweight, efficient fusion of retrieved documents during decoding, yielding more moderate improvements of 18--27\% across TriviaQA~\cite{joshi-etal-2017-triviaqa} and NQ~\cite{kwiatkowski-etal-2019-natural}. R$^2$AG, another generator-based approach, shows promising gains, with over 80\% improvement from the baseline on NQ, further validating the benefits of integrating retrieval signals within generation. We note that generator-based frameworks primarily designed for efficiency, such as xRAG, are discussed separately due to their distinct optimization focus.

Retriever-based frameworks such as RQ-RAG and SimRAG also demonstrate notable gains. RQ-RAG achieves a 288\% improvement on PopQA and over 210\% on ARC-Challenge, reaffirming the importance of retrieval quality in evidence-centric QA. SimRAG also shows strong improvements on ARC, although gains are more modest (approximately 14\%). Additionally, retriever-side re-ranking approaches like FILCO deliver moderate but meaningful gains, with 5--30\% improvements across NQ and TriviaQA, further highlighting the incremental value of retrieval refinement strategies.

Hybrid frameworks exhibit a more heterogeneous pattern. CRAG and Self-CRAG achieve impressive gains, with Self-CRAG delivering a 320\% improvement on PopQA and a 208\% improvement on ARC-Challenge, suggesting that combining retrieval refinement with generation adaptation can be highly effective when well aligned. However, TA-ARE, despite achieving a significant 28$\times$ improvement over raw baselines on RetrievalQA, occasionally underperforms relative to the standard retrieval baseline, indicating that retrieval frequency reduction strategies, while efficient, may introduce trade-offs. Stochastic RAG frameworks, meanwhile, display relatively modest gains (typically under 4\%), reflecting that introducing retrieval randomness increases diversity without consistently boosting short-form QA accuracy.

Efficiency-focused generator-based systems such as xRAG exhibit mixed results. While xRAG achieves 10--29\% improvements over raw LLM baselines on datasets such as NQ and TriviaQA, its gains over retrieval baselines are marginal or occasionally negative. This suggests that while resource-efficient designs are promising for scaling RAG systems, further optimization is needed to maintain competitive factual accuracy in short-form tasks.

Finally, robustness-oriented frameworks such as RAAT demonstrate strong performance, with a 116\% improvement from the raw baseline and over 27\% gain compared to retrieval on RAG-Bench---a robustness-focused variant of NQ, WebQ, and TriviaQA. Although evaluated under challenging retrieval noise conditions, RAAT’s results suggest that robustness-driven retrieval strategies can effectively complement factual QA objectives.

Overall, retrieval- and generation-enhanced frameworks deliver substantial relative gains in short-form QA, while hybrid and efficiency-focused approaches offer promising but variable results depending on dataset and retrieval complexity. These findings underscore the critical role of retrieval optimization and generation-adaptive strategies in advancing retrieval-augmented short-form question answering.

\subsection{Comparative Analysis of Framework Performance on Multi-Hop QA}

A comparative evaluation of various Retrieval-Augmented Generation (RAG) frameworks reveals distinct patterns in their ability to enhance multi-hop question answering, assessed through improvements over both raw large language models (LLMs) and standard retrieval-augmented baselines. Similar to the previous section, this analysis focuses on relative gains rather than absolute scores to normalize for architectural and experimental variations. The results, summarized in Table~\ref{tab:comparativeanalysis}, enable a consistent comparison of framework contributions across diverse multi-hop QA settings.

Among retrieval-based RAG systems, models such as RQ-RAG, RankRAG, LQR, and LongRAG demonstrate substantial relative gains. Notably, RQ-RAG achieves over an 800\% improvement from its raw LLM baseline on HotpotQA~\cite{yang-etal-2018-hotpotqa}, and a 275\% improvement over standard retrieval, highlighting the effectiveness of sophisticated query decomposition techniques in multi-hop settings. Similarly, LQR achieves a 292\% improvement from the raw baseline and an 84\% improvement over retrieval in the MuSiQue dataset~\cite{trivedi-etal-2022-musique}, suggesting that intelligent retrieval-ranking substantially boosts multi-hop reasoning. LongRAG also exhibits strong performance, improving by over 50\% from the raw LLM baseline on HotpotQA, further emphasizing the value of extended retrieval for complex question answering. These patterns collectively affirm that optimizing retrieval quality remains a dominant driver of performance gains in multi-hop RAG applications.

Generator-based RAG frameworks, including R$^2$AG, INFO-RAG, and xRAG, display more varied relative improvements. R$^2$AG shows consistent strong gains, improving by over 300\% relative to the baseline on HotpotQA, demonstrating the benefits of tightly integrating retrieval signals into the generation process. In contrast, INFO-RAG exhibits more modest improvements, with relative gains around 16--35\% across different backbones and datasets, suggesting that while generator-side augmentations enhance output faithfulness, their standalone effect may be limited without concurrent retrieval refinement. xRAG, while improving from raw baselines by approximately 20--26\%, shows negative or marginal gains compared to the retrieval baseline in some settings, indicating that extreme context compression, although efficient, may compromise the model’s ability to utilize retrieved evidence effectively for complex multi-hop reasoning.

Hybrid RAG frameworks, such as DRAGIN, FLARE, GenGround, and Stochastic RAG, present a diverse range of outcomes. DRAGIN frameworks achieve moderate improvements, typically ranging between 22--44\% over raw LLMs and 14--34\% over retrieval baselines, reflecting the incremental gains from dynamically adapting retrieval to evolving information needs. FLAREdirect stands out, achieving a 62\% improvement from the raw LLM and a 22\% improvement over standard retrieval on 2Wiki~\cite{ho-etal-2020-constructing}, suggesting that model-guided active retrieval significantly strengthens multi-hop evidence gathering. GenGround reports relatively smaller improvements (13--36\% from the baseline) but is evaluated against already-strong baselines, which partially accounts for the more conservative gains. Stochastic RAG frameworks offer consistent yet modest gains (6\%), indicating that introducing randomness into retrieval can modestly diversify and enhance evidence coverage without destabilizing performance.

Overall, retrieval-based RAG frameworks demonstrate the most consistent and substantial improvements across multi-hop QA tasks, particularly when retrieval quality, ranking, and query decomposition are optimized. Generator-based adaptations, while beneficial in specific cases, often require complementary retrieval-side enhancements to realize their full potential. Hybrid frameworks offer promising but more variable results, underscoring the challenge of harmonizing retrieval and generation strategies dynamically. These findings highlight retrieval optimization as a critical lever for advancing complex reasoning capabilities in RAG systems.

\subsection{Comparative Robustness Analysis: Framework Gains Over Retrieval-Only Baselines}

To assess robustness in Retrieval-Augmented Generation (RAG) systems, we report incremental improvements each framework achieves over its retrieval-augmented LLM baseline. This isolates the added value of mechanisms such as critique, reranking, and filtering, independent of the baseline retrieval gain. Evaluations span multiple datasets and focus on gains in precision, recall, and FactScore. By standardizing on relative improvements, the analysis enables fair comparisons across models with differing backbone architectures. A summary of these results is provided in Table~\ref{tab:robustness}.

Among hybrid systems, the most substantial gains in factual consistency are observed. Self-CRAG yields the highest FactScore improvement---+0.456 on the Biography dataset~\cite{min-etal-2023-factscore}---significantly surpassing other frameworks, most of which report $\leq$0.05 gains. The multi-sentence compositional nature of the Biography task likely benefits from Self-CRAG’s feedback-based reranking and correction loop, which aligns generation with retrieved evidence. Comparable improvements are evident with Self-RAG and CRAG, reporting +0.372 and +0.252 gains on the same dataset, underscoring the importance of evidence-aware generation refinement. On 2Wiki, Flare-Direct improves both precision and recall by +21.6\%, while Flare-Instruct---despite using the same retrieval backbone---offers negligible gains, illustrating how prompt design alone can meaningfully impact robustness in multi-hop settings. In contrast, Stochastic RAG shows only marginal FactScore gains ($\leq$+0.008) on Fever, suggesting that entropy-driven retrieval without subsequent verification may be insufficient to ensure factual reliability.

Generator-based systems present more variable, task-dependent performance. SELF-RAG, evaluated on ASQA~\cite{stelmakh-etal-2022-asqa}, achieves sizable improvements in precision (+22--30\%) and recall (+16--19\%), though its FactScore gains remain modest (+0.03--0.04), implying improved evidence usage without equivalent advances in factual accuracy. DRAGIN similarly improves precision and recall by +9--22\% on HotPotQA, leveraging entropy-based token-level triggers suited for multi-hop reasoning. However, lacking reported FactScore, its contribution to factual consistency remains indeterminate. Other generator-oriented systems, including GenRT and Rich Answer Encoding, achieve smaller recall gains ($\leq$+0.1) on datasets such as TriviaQA, KILT-WoW~\cite{petroni-etal-2021-kilt}, and MSMARCO~\cite{bajaj2018msmarcohumangenerated}. These modest improvements suggest better document selection but limited post-retrieval validation, constraining their robustness impact.

Retriever-based systems exhibit consistent yet comparatively modest gains. Re2G reports +17.8\% precision and +15.9\% recall on TriviaQA, reflecting the benefits of retrieval-aware prompt optimization. FILCO, by contrast, improves precision by +3.25\% on Fever but fails to enhance recall or FactScore, indicating that filtering irrelevant context improves selectivity, but without downstream verification, its robustness contribution is limited. Not all frameworks report all three metrics across datasets; while relative improvement facilitates normalization, incomplete coverage---particularly of FactScore---may obscure the full extent of a system’s capabilities.

In sum, Self-CRAG on Biography delivers the strongest FactScore gain (+0.456), while SELF-RAG on ASQA achieves the best precision (+29.56\%) and recall (+18.81\%) improvements. Flare-Direct, outperforming Flare-Instruct by over 20\% on 2Wiki, highlights the sensitivity of robustness to prompt design. At the lower end, Stochastic RAG on FEVER~\cite{thorne-etal-2018-fever} records the smallest impact ($\leq$+0.008 FactScore), reinforcing the necessity of combining retrieval strategies with downstream verification to enhance factual fidelity.

Collectively, these findings affirm that retrieval alone is insufficient for robust generation. The most effective frameworks tightly couple retrieval, generation, and verification in iterative loops, ensuring that generation is guided by critique and alignment rather than treated as a terminal step.

\begin{table*}[t]
\centering
\caption{\textbf{Comparative Robustness Analysis of RAG Frameworks Across Architectures.} Relative improvements in precision, recall, and FactScore over retrieval-augmented baselines across multiple datasets. A dash (--) denotes missing values in the original paper.}
\label{tab:robustness}
\scriptsize
\begin{tabular}{llllccc}
\toprule
\textbf{Taxonomy} & \textbf{Framework} & \textbf{Backbone} & \textbf{Dataset} & \textbf{Precision} & \textbf{Recall} & \textbf{FactScore} \\
\midrule
\multirow{4}{*}{Retriever-based RAG} 
& Re2G & KGI0 & NQ & 0.096984 & 0.074569 & -- \\
& Re2G & KGI1 & TriviaQA & 0.177981 & 0.159062 & -- \\
& Re2G & KGI2 & Fever & 0.120986 & 0.073732 & -- \\
& FILCO & RAG & Fever & 3.25 & -- & -- \\
\midrule
\multirow{8}{*}{Generation-based RAG}
& SELF-RAG & LLaMA2-7B & ASQA & 22.06897 & 15.95 & 0.041026 \\
& SELF-RAG & LLaMA2-7B & ASQA & 29.56522 & 18.80556 & 0.034839 \\
& Rich Answer Encoding & RAG & MSMARCO & -- & 0.086957 & -- \\
& Rich Answer Encoding & RAG & KILT-WoW & -- & 0.107293 & -- \\
& DRAGIN & LLaMA2-13B & HotPotQA & 0.185934 & 0.09893 & -- \\
& DRAGIN & VICUNA-13B & HotPotQA & 0.222447 & 0.105114 & -- \\
& GenRT & RAG & NQ & -- & 0.023232 & -- \\
& GenRT & RAG & TriviaQA & -- & 0.026239 & -- \\
\midrule
\multirow{6}{*}{Hybrid RAG}
& CRAG & LLaMA2-7B & Biography & -- & -- & 0.251689 \\
& Self-CRAG & LLaMA2-7B & Biography & -- & -- & 0.456081 \\
& Flare-Instruct & GPT-3.5 & 2Wiki & 0.010288 & 0.019417 & -- \\
& Flare-Direct & GPT-3.5 & 2Wiki & 0.216049 & 0.215534 & -- \\
& Stochastic RAG & FiD-Light (T5-Base) & Fever & -- & -- & 0.008685 \\
& Stochastic RAG & FiD-Light (T5-XL) & Fever & -- & -- & 0.00355 \\
\bottomrule
\end{tabular}
\end{table*}

\subsection{Ablation Studies}

Ablation studies serve as a crucial methodological lens for disentangling the contributions of individual components in Retrieval-Augmented Generation (RAG) frameworks. Across the surveyed literature, these studies primarily target retrieval triggers, filtering layers, reranking strategies, compression modules, and corrective mechanisms, offering empirical insights into performance, efficiency, and robustness.

\textbf{Adaptive Retrieval and Query Reformulation.} Frameworks such as TA-ARE, FLARE, and IM-RAG demonstrate that retrieval adaptivity is central to long-form and multi-hop reasoning. Ablating dynamic query triggers (e.g., forward-looking or self-reflective prompts) consistently results in degraded factual accuracy and increased hallucinations, confirming the value of retrieval-awareness throughout the generation process.

\textbf{Filtering, Reranking, and Evidence Quality.} Systems like SEER, CRAG, and Re2G show that context filtering, reranking, and correction layers significantly influence downstream performance. Ablations reveal that removing context evaluators or decomposing mechanisms leads to verbosity and reduced grounding fidelity. Notably, reranking-truncation co-designs (e.g., in GenRT and ToolRerank) outperform static top-$k$ approaches by improving answer faithfulness and retrieval precision.

\textbf{Compression and Efficiency Trade-offs.} FiD-Light and RAGCache demonstrate that passage compression and caching can substantially reduce latency without compromising accuracy. Ablating vector sparsity or caching mechanisms (e.g., speculative pipelining or prefix-aware replacement) increases inference time up to 4$\times$, underscoring the operational significance of architectural optimization in production RAG systems.

\textbf{Robustness and Security.} Studies like BadRAG and TrojanRAG emphasize that security-focused ablations reveal novel vulnerabilities. Even lightweight retrieval poisoning or trigger crafting can steer model outputs, while mitigation strategies (e.g., summarization, distance thresholds) offer partial resilience but require further study.

\textbf{Synthesis.} Ablation studies consistently reinforce that high-performing RAG frameworks are modular, with complementary retrieval, filtering, and generation components. Performance degradation in ablation settings not only validates novel modules but also guides design toward more interpretable, efficient, and secure RAG pipelines.

\section{Evaluation and Benchmarking of RAG Systems}
\label{sec:evaluation}
Retrieval-Augmented Generation (RAG) systems introduce unique challenges for evaluation due to their hybrid architecture combining a retriever and a generator. Accurate evaluation demands assessing multiple interdependent components, including retrieval relevance, faithfulness of generated responses, and overall answer utility. In this section, we synthesize recent advancements in automated evaluation frameworks, retrieval quality assessment techniques, and benchmark construction to provide a comprehensive overview of evaluation practices in RAG systems.

\subsection{Evaluation Dimensions}

The core dimensions ~\cite{saad-falcon-etal-2024-ares} used to evaluate RAG systems include:
\begin{enumerate}
    \item \textbf{Context Relevance:} Measures how pertinent the retrieved documents are to the input query.
    \item \textbf{Answer Faithfulness:} Assesses whether the generated output remains grounded in the retrieved evidence.
    \item \textbf{Answer Relevance:} Evaluates whether the output adequately addresses the user query.
\end{enumerate}

These dimensions are interdependent: poor context relevance often cascades into reduced faithfulness and answer relevance, underscoring the need for joint evaluation. Frameworks such as ARES and RAGAS have formalized these dimensions, incorporating both automated judgment and reference-free evaluation.

\subsection{Automated Evaluation Frameworks}

ARES~\cite{saad-falcon-etal-2024-ares} introduces an LLM-based judge system that uses few-shot prompted language models to generate synthetic datasets. These judges are trained on three classification tasks corresponding to the core dimensions and use prediction-powered inference (PPI) to align model-based scoring with human judgment. ARES shows significant improvements in accuracy and annotation efficiency, outperforming RAGAS~\cite{es-etal-2024-ragas} by up to 59.3 percentage points in context relevance.

RAGAS employs a modular framework that decomposes generated answers into atomic factual statements, then evaluates each against the retrieved context using LLMs. This structure provides high-resolution feedback, revealing which parts of an answer are hallucinated.

These frameworks automate the evaluation of faithfulness, grounding, and contextual relevance---enabling scalable, reference-free analysis of RAG performance.

\subsection{Evaluating Retrieval Quality}

eRAG~\cite{salemi2024erag} challenges traditional relevance label techniques by applying the RAG generator to each retrieved document individually. The performance of each document, assessed via downstream task metrics, serves as a relevance label. This method provides a retrieval-aware, document-level granularity and has shown significantly improved correlation with actual RAG performance.

INFO-RAG introduces an unsupervised training paradigm that improves the LLM’s ability to refine retrieved information under three scenarios: redundant, noisy, or insufficient context. By viewing the LLM as an “information refiner,” it enables the model to extract relevant content, reject misinformation, and infer missing details---enhancing retrieval robustness without supervised relevance labels.

uRAG proposes a unified retrieval system that serves multiple RAG models across diverse downstream tasks. It introduces a shared reranker trained on feedback signals (e.g., EM, accuracy) from various black-box LLMs, treating each LLM as a user of the search engine. uRAG’s training protocol enables evaluation and optimization of retrieval based on downstream task performance, offering retrieval diagnostics grounded in actual utility rather than surface similarity.

\subsection{Benchmarking RAG Capabilities}

As RAG systems mature, a growing suite of benchmarks has emerged to evaluate them across dimensions like robustness, factuality, adaptivity, and domain sensitivity. These benchmarks not only reflect the evolving needs of real-world RAG deployments but also shape future directions by surfacing recurrent failure modes and task-specific limitations.

\textbf{Robustness to retrieval noise} is a core requirement in operational RAG systems. RGB~\cite{chen2024rgb} evaluates four fundamental capacities—noise robustness, negative rejection, information integration, and counterfactual resistance—revealing consistent weaknesses in LLMs when handling distracting or misleading context. Complementing this, RAG-Bench~\cite{fang-etal-2024-enhancing} introduces a noise-centric benchmark simulating three retrieval corruption types—relevant-but-incomplete, irrelevant, and counterfactual—and applies adaptive adversarial training to improve model tolerance. These benchmarks enable fine-grained analysis of how retrieval perturbations degrade end-task performance and inform robust retrieval-policy design.

\textbf{Faithfulness and hallucination detection} benchmarks have taken center stage in evaluating generation quality. RAGTruth~\cite{niu-etal-2024-ragtruth} provides nearly 18{,}000 annotated examples from QA, summarization, and data-to-text generation, offering both response- and span-level hallucination labels across four types: subtle vs. evident, and conflict vs. baseless information. Uniquely, it supports training hallucination detectors and benchmarking span-level detection precision and recall—tasks not addressed by coarse-grained metrics. This makes it foundational for measuring factual integrity in RAG outputs.

\textbf{Reasoning and retrieval chaining} are central to multi-hop question answering, where evidence spans multiple documents. MultiHop-RAG~\cite{tang2024multihoprag} targets this challenge through linked question-answer pairs, bridge entities, and explicit multi-hop query types, enabling systematic assessment of retrieval chaining, evidence linking, and document-level reasoning—all key bottlenecks in complex RAG workflows.

\textbf{Adaptive retrieval and necessity estimation} are benchmarked in RetrievalQA~\cite{zhang-etal-2024-retrievalqa}, which mixes queries requiring external retrieval with those answerable via the base LLM alone. This design tests whether models can intelligently toggle retrieval based on query uncertainty, supporting the development of resource-efficient, retrieval-aware systems that avoid introducing unnecessary context.

\textbf{Domain-specific evaluation} is exemplified by MIRAGE~\cite{confcalrag2025}, a benchmark tailored to medical RAG. It contains 7{,}663 questions sourced from five clinical and biomedical QA datasets and incorporates real-world evaluation constraints: zero-shot generalization, multiple-choice formats, retrieval necessity assessment, and question-only retrieval. This multi-faceted setup tests reliability under high-stakes conditions where factual errors can be consequential.

\textbf{Cross-corpus and federated retrieval} are explored in FeB4RAG~\cite{wang2024feb4rag}, a benchmark constructed from 16 BEIR sub-collections. It evaluates federated retrieval through 790 conversational queries with LLM-graded relevance judgments and quantifies the impact of resource selection and result merging strategies. This benchmark surfaces key risks in multi-source RAG pipelines, especially retrieval inconsistency and hallucination amplification due to poor corpus coordination.

\textbf{Evaluation infrastructure and reproducibility} are addressed by BERGEN~\cite{rau-etal-2024-bergen}, a benchmarking library designed to unify assessment across RAG components. It offers modular templates for measuring retrieval precision, generation faithfulness, and their interplay across datasets and model configurations. BERGEN facilitates consistent and extensible RAG benchmarking in both academic and applied settings.

\begin{table*}[t]
\centering
\caption{\textbf{Emerging Benchmarks for Evaluating Retrieval-Augmented Generation (RAG) Systems.} This table summarizes recent benchmarks developed to assess key aspects of RAG systems, including robustness, multi-hop reasoning, medical-domain adaptation, and federated retrieval. These benchmarks differ in evaluation granularity—ranging from query-level to document-level—and employ varied annotation methods such as manual labeling, programmatic perturbation, and LLM-based scoring. Distinctive features, such as noise stress-testing (RGB), zero-shot medical QA (MIRAGE), and federated source merging (FeB4RAG), support targeted evaluations of both retriever components and full RAG pipelines.}
\label{tab:rag-benchmarks}
\resizebox{\textwidth}{!}{%
\begin{tabular}{p{2.2cm} p{2.9cm} p{2.2cm} p{2.2cm} p{3.4cm} p{2.2cm}}
\toprule
\textbf{Benchmark} & \textbf{Evaluation Focus} & \textbf{Granularity} & \textbf{Annotation Type} & \textbf{Unique Features} & \textbf{Evaluation Target} \\
\midrule
RGB & Robustness (noise, integration, hallucination) & Query-context pair & None & Stress tests for noise, contradiction, and multi-source fusion & Full pipeline \\
MultiHop-RAG & Multi-hop reasoning and retrieval chaining & Document-level & Manual + derived & Linked multi-hop queries and bridge-entity chaining & Full pipeline \\
RAGTruth & Hallucination detection and factuality evaluation & Response-level (yes/no), span-level (exact) & Human-labeled & 18,000+ examples, 4 hallucination types, span-level F1 & Generator \\
MIRAGE & Medical domain QA under real-world constraints & Query-level & Dataset-native & Zero-shot, multi-choice, question-only retrieval (MEDRAG) & Full pipeline \\
FeB4RAG & Federated retrieval evaluation & Document + resource & LLM-labeled & Measures retrieval + merging across 16 BEIR sources & Retriever \\
RetrievalQA & Adaptive retrieval necessity detection & Query-level & Derived & Queries with and without need for retrieval & Retriever \\
RAG-Bench & Retrieval robustness to noise & Query-level & Programmatic & Irrelevant, incomplete, and counterfactual retrieval noise & Full pipeline \\
BERGEN & Retrieval, generation, and joint evaluation & Query-context and document-level & Configurable (task-dependent) & Unified benchmarking library across datasets and models & Full pipeline \\
\bottomrule
\end{tabular}
}
\end{table*}

This section outlines the rapidly evolving landscape of RAG evaluation and benchmarking. Future RAG development hinges not only on improving generation quality but also on designing principled, scalable, and interpretable evaluation strategies that reflect real-world usage and complexities. To advance the field meaningfully, the community must prioritize the creation of standardized, efficient, and adaptive evaluation protocols that can serve both research and production contexts.

\section{Future Directions}
\label{sec:future}
As Retrieval-Augmented Generation (RAG) systems continue to evolve, a number of unresolved challenges remain that limit their deployment in dynamic, open-ended, and high-stakes applications. These challenges span retrieval efficiency, semantic misalignment, hallucination control, generalization, and trust. Based on the synthesis of contemporary research gaps, we outline five interrelated future directions that represent promising trajectories for advancing the field.

\subsection{Retrieval Adaptivity and Semantic Alignment}

Current RAG architectures often rely on static retrieval policies and fixed embedding transformations, limiting their adaptability to complex or evolving user queries. Future systems must support dynamically calibrated retrieval strategies that adjust depth, modality, and source selection in response to task difficulty and contextual cues. This calls for co-optimized retriever–generator pipelines that leverage reinforcement signals, uncertainty estimates, or semantic control layers to align evidence retrieval with generative intent in real time.

\subsection{Robustness under Noise and Adversarial Conditions}

Despite recent advances in noise filtering and adversarial training, RAG systems remain vulnerable to retrieval perturbations, misleading content, and corpus-level poisoning attacks. Future work should move toward retrieval-aware adversarial defenses that incorporate noise-aware loss functions, retrieval-type-specific regularization, and semantic provenance filtering. This includes evaluation protocols that stress-test systems against contextually plausible yet misleading passages and group-triggered semantic attacks, as exemplified by recent backdoor threat models.

\subsection{Multi-Hop Reasoning and Structured Compositionality}

Many knowledge-intensive tasks require aggregating evidence across multiple retrieval steps and reasoning over entity or schema-level structures. Current models exhibit limited capacity for compositional inference or procedural synthesis. Future RAG systems should support multi-turn retrieval–generation loops, structured subgoal decomposition, and graph-augmented reasoning pipelines that maintain discourse coherence and entity consistency across long-range dependencies.

\subsection{Cross-Domain Generalization and Temporal Adaptivity}

RAG performance often degrades in the face of domain shifts, novel schema, or temporal drift. Addressing this will require pretraining retrieval modules on diverse proxy tasks, developing meta-retrievers capable of adapting to unseen query distributions, and incorporating recency-aware document scoring. Additionally, the design of temporally evolving benchmarks and evaluation suites will be necessary to assess the robustness of RAG systems under realistic, time-sensitive knowledge conditions.

\subsection{Explainability, Personalization, and Trust Calibration}

As RAG systems are increasingly integrated into user-facing applications, demands for interpretability, personalization, and secure behavior intensify. Future architectures should expose transparent interfaces for explaining retrieval decisions and generation provenance, while supporting privacy-preserving personalization through user-clustered retrieval, memory-efficient modeling, or differential privacy mechanisms. Furthermore, integrating retrieval calibration signals—such as factual salience, source trustworthiness, or hallucination risk—can enhance user trust and system accountability.

\bibliographystyle{ACM-Reference-Format}
\bibliography{references}

\appendix
\section{Appendix}

To support transparency and reproducibility, we include in the Appendix the original benchmark scores reported in the primary publications of each RAG framework. These tables serve as the empirical source for the relative improvement analyses presented in Sections~\ref{sec:comparison}. All values are cited from original papers, preserving reported metrics such as F1, EM, Accuracy, and FactScore. Where applicable, dataset splits, backbone models, and evaluation metrics are clearly labeled to ensure traceability.

\clearpage

\begin{table*}[t]
\centering
\caption{\textbf{Reported Performance Scores for Short-Form QA Frameworks.} Accuracy and Exact Match (EM) scores as reported in the original publications of short-form RAG frameworks. These values were used to compute the normalized improvements presented in Section~\ref{sec:comparison}.}
\label{tab:appendix-shortform}
\scriptsize
\resizebox{\textwidth}{!}{%
\begin{tabularx}{\textwidth}{l l l l c c c c}
\toprule
\textbf{Taxonomy} & \textbf{Framework} & \textbf{Backbone} & \textbf{Dataset} & \textbf{Metric} & \textbf{Raw LLM} & \textbf{LLM+Retrieval} & \textbf{Framework Score} \\
\midrule
\multirow{8}{*}{Retriever-Based RAG}
& RQ-RAG & LLaMA2-7B & PopQA & Acc & 14.7 & 39.8 & 57.1 \\
& RQ-RAG & LLaMA2-7B & ARC-Challenge & Acc & 21.8 & 28.7 & 68.3 \\
& SimRAG & LLaMA3-8B & ARC-Challenge & Acc & -- & 71.08 & 81.4 \\
& SimRAG & LLaMA3-8B & SciQ & EM & -- & 20.8 & 57.5 \\
& SimRAG & Gemma2-27B & ARC-Challenge & Acc & -- & 85.75 & 88.65 \\
& SimRAG & Gemma2-27B & SciQ & EM & -- & 44.8 & 58.1 \\
& Re2G & BART Large & NQ & Acc & 45.22 & -- & 51.73 \\
& Re2G & BART Large & TriviaQA & Acc & 60.99 & -- & 76.27 \\
& FILCO & LLaMA2-7B (Top-5) & NQ & EM & -- & 47.6 & 61.8 \\
& FILCO & LLaMA2-7B (Top-5) & TriviaQA & EM & -- & 67.3 & 71.1 \\
\midrule
\multirow{11}{*}{Generator-Based RAG}
& SELF-RAG & LLaMA2-7B & PopQA & Acc & 14.7 & 38.2 & 54.9 \\
& SELF-RAG & LLaMA2-7B & TriviaQA & Acc & 30.5 & 42.5 & 66.4 \\
& SELF-RAG & LLaMA2-7B & ARC-Challenge & Acc & 21.8 & 48.0 & 67.4 \\
& SELF-RAG & LLaMA2-13B & PopQA & Acc & 14.7 & 45.7 & 55.8 \\
& SELF-RAG & LLaMA2-13B & TriviaQA & Acc & 38.5 & 47.0 & 69.3 \\
& xRAG & Mistral-7B & NQ & EM & 30.25 & 42.71 & 39.1 \\
& xRAG & Mistral-7B & TriviaQA & EM & 57.08 & 65.88 & 65.77 \\
& xRAG & Mistral-7B & WebQA & EM & 34.89 & 37.84 & 39.4 \\
& xRAG & Mixtral-8x7B & NQ & EM & 41.99 & 45.15 & 47.28 \\
& xRAG & Mixtral-8x7B & TriviaQA & EM & 71.1 & 70.34 & 74.14 \\
& xRAG & Mixtral-8x7B & WebQA & EM & 40.31 & 41.26 & 44.5 \\
& FiD-Light & FiD+DPR & TriviaQA & EM & 48.6 & -- & 57.6 \\
& FiD-Light & FiD+DPR & NQ & EM & 41.9 & -- & 53.2 \\
& R$^2$AG & LLaMA2-7B & NQ & Acc & 0.38 & -- & 0.693 \\
& SELF-RAG & LLaMA2-7B & NQ & Acc & 0.38 & -- & 0.188 \\
\midrule
\multirow{11}{*}{Hybrid RAG}
& Stochastic RAG & FiD-Light (T5-Base) & NQ & EM & -- & 45.6 & 46.2 \\
& Stochastic RAG & FiD-Light (T5-Base) & TriviaQA & EM & -- & 57.6 & 59.7 \\
& Stochastic RAG & FiD-Light (T5-XL) & NQ & EM & -- & 51.1 & 53.0 \\
& Stochastic RAG & FiD-Light (T5-XL) & TriviaQA & EM & -- & 63.7 & 64.7 \\
& CRAG & LLaMA2-7B & NQ & Acc & 0.38 & -- & 0.397 \\
& CRAG & LLaMA2-7B & PopQA & Acc & 14.7 & 40.3 & 59.3 \\
& CRAG & LLaMA2-7B & ARC-Challenge & Acc & 21.8 & 46.7 & 54.8 \\
& Self-CRAG & LLaMA2-7B & PopQA & Acc & 14.7 & 40.3 & 61.8 \\
& Self-CRAG & LLaMA2-7B & ARC-Challenge & Acc & 21.8 & 46.7 & 67.2 \\
& TA-ARE & GPT-3.5 & RetrievalQA & Acc & 1.2 & 38.2 & 35.8 \\
& TA-ARE & GPT-4 & RetrievalQA & Acc & 2.4 & 46.0 & 46.4 \\
& TA-ARE & LLaMA2-7B & RetrievalQA & Acc & 2.0 & 36.0 & 30.7 \\
\midrule
\multirow{1}{*}{Robustness-Based RAG}
& RAAT & LLaMA2-7B & RAG-Bench (TQA/NQ/WebQ) & EM & 38.37 & 65.4 & 83.07 \\
\bottomrule
\end{tabularx}
}
\end{table*}

\begin{table*}[t]
\centering
\caption{\textbf{Reported Performance Scores for Multi-Hop QA Frameworks.} Raw F1 and EM scores extracted from the original papers of multi-hop RAG systems, across datasets such as HotpotQA, 2Wiki, and MuSiQue. These scores form the basis of the comparative analysis in Section~\ref{sec:comparison}.}
\label{tab:appendix-multihop}
\small
\resizebox{\textwidth}{!}{%
\begin{tabular}{lllp{2.2cm}cccc}
\toprule
\textbf{Taxonomy} & \textbf{Framework} & \textbf{Backbone} & \textbf{Dataset} & \textbf{Metric} & \textbf{Raw LLM} & \textbf{LLM + Retrieval} & \textbf{Framework Score} \\
\midrule
\multirow{11}{*}{Retriever-Based} 
& RQ-RAG & LLaMA2-7B & HotpotQA & F1 & 6.6 & 16.7 & 62.6 \\
& RQ-RAG & LLaMA2-7B & 2Wiki & F1 & 16 & 18.7 & 44.8 \\
& RQ-RAG & LLaMA2-7B & MuSiQue & F1 & 3 & 7.4 & 41.7 \\
& RankRAG & LLaMA3-8B & HotpotQA & F1 & -- & 43.3 & 46.7 \\
& RankRAG & LLaMA3-8B & 2Wiki & F1 & -- & 27.9 & 36.9 \\
& RankRAG & LLaMA3-70B & HotpotQA & F1 & -- & 44.6 & 55.4 \\
& RankRAG & LLaMA3-70B & 2Wiki & F1 & -- & 31.9 & 43.9 \\
& LQR & LLaMA3-8B & MuSiQue & F1 & 10.7 & 22.8 & 41.97 \\
& LQR & LLaMA3-8B & HotpotQA & F1 & 22.71 & 46.15 & 69.96 \\
& LQR & LLaMA3-8B & 2Wiki & F1 & 32.04 & 47.9 & 54.65 \\
& LongRAG & GPT-4o & HotpotQA & EM & 42.4 & -- & 64.3 \\
& LongRAG & Gemini-1.5-Pro & HotpotQA & EM & 33.9 & -- & 57.5 \\
& SEER & LLaMA2-7B-Chat & HotpotQA & F1 & 0.5471 & 0.5826 & 0.604 \\
& FILCO & LLaMA2-7B & HotpotQA & EM & -- & 61.5 & 65 \\
\midrule
\multirow{10}{*}{Generator-Based}
& R$^2$AG & LLaMA2-7B & HotpotQA & F1 & 8.52 & -- & 36.05 \\
& R$^2$AG & LLaMA2-7B & MuSiQue & F1 & 2.41 & -- & 16.87 \\
& R$^2$AG & LLaMA2-7B & 2Wiki & F1 & 6.34 & -- & 34.52 \\
& xRAG & Mistral-7B & HotpotQA & EM & 27.02 & 38.79 & 34.05 \\
& xRAG & Mixtral-8x7B & HotpotQA & EM & 32.87 & 43.46 & 39.66 \\
& INFO-RAG & LLaMA2-7B & HotpotQA & EM & 39.4 & -- & 46.56 \\
& INFO-RAG & LLaMA2-13B & HotpotQA & EM & 42.12 & -- & 51.48 \\
& INFO-RAG & LLaMA2-13B-chat & HotpotQA & EM & 61.23 & -- & 61.91 \\
& INFO-RAG & LLaMA2-7B & MuSiQue & EM & 25.95 & -- & 30.19 \\
& INFO-RAG & LLaMA2-13B & MuSiQue & EM & 25.78 & -- & 35.02 \\
& INFO-RAG & LLaMA2-13B-chat & MuSiQue & EM & 47.06 & -- & 47.93 \\
\midrule
\multirow{13}{*}{Hybrid}
& DRAGIN & LLaMA2-13B-chat & HotpotQA & F1 & 30.97 & 37.06 & 42.38 \\
& DRAGIN & LLaMA2-13B-chat & 2Wiki & F1 & 27.21 & 33.64 & 39.31 \\
& DRAGIN & LLaMA2-7B-chat & HotpotQA & F1 & 27.45 & 24.99 & 33.44 \\
& DRAGIN & LLaMA2-7B-chat & 2Wiki & F1 & 22.32 & 25.49 & 29.26 \\
& DRAGIN & Vicuna-13B-v1.5 & HotpotQA & F1 & 32.56 & 35.31 & 41.64 \\
& DRAGIN & Vicuna-13B-v1.5 & 2Wiki & F1 & 22.32 & 25.64 & 35.16 \\
& FLAREdirect & GPT-3.5 & 2Wiki & F1 & 36.8 & 48.8 & 59.7 \\
& FLAREinstruct & GPT-3.5 & 2Wiki & F1 & 36.8 & 48.8 & 49.8 \\
& GenGround & GPT-3.5 & HotpotQA & F1 & 42.28 & 47.8 & 52.26 \\
& GenGround & GPT-3.5 & MuSiQue & F1 & 20.13 & 20.11 & 27.36 \\
& GenGround & GPT-3.5 & 2Wiki & F1 & 41.19 & 44.77 & 50.21 \\
& GenGround & GPT-3.5 & StrategyQA & F1 & 68.13 & 71.78 & 77.12 \\
& Stochastic RAG & FiD-Light (T5-Base) & HotpotQA & EM & 25.6 & -- & 27.3 \\
& Stochastic RAG & FiD-Light (T5-XL) & HotpotQA & EM & 29.2 & -- & 31.1 \\
\bottomrule
\end{tabular}
}
\end{table*}

\begin{table*}[t]
\centering
\caption{\textbf{Reported Robustness Scores for RAG Frameworks.} Precision, recall, and FactScore values extracted from original publications, across multiple datasets. These scores serve as the empirical basis for the comparative robustness analysis in Section~\ref{sec:comparison}.}
\label{tab:appendix-robustness}
\small
\resizebox{\textwidth}{!}{%
\begin{tabular}{lllp{3.2cm}ccc}
\toprule
\textbf{Taxonomy} & \textbf{Framework} & \textbf{Backbone} & \textbf{Metric (Dataset)} & \textbf{LLM + Retrieval} & \textbf{LLM + Retrieval + Framework} \\
\midrule
\multirow{7}{*}{Retriever-Based RAG}
& FILCO & RAG & Precision (FEVER) & 1.2 & 5.1 \\
& Re2G & KGI0 & Precision (NQ) & 64.65 & 70.92 \\
& Re2G & KGI0 & Recall (NQ) & 69.6 & 74.79 \\
& Re2G & KGI0 & R-Precision (NQ) & 61.13 & 72.01 \\
& Re2G & KGI0 & Recall (TriviaQA) & 63.12 & 73.16 \\
& Re2G & KGI0 & R-Precision (FEVER) & 80.34 & 90.06 \\
& Re2G & KGI0 & Recall (FEVER) & 86.53 & 92.91 \\
\midrule
\multirow{10}{*}{Generator-Based RAG}
& SELF-RAG & LLaMA2-7B & Precision (ASQA) & 2.9 & 66.9 \\
& SELF-RAG & LLaMA2-7B & Recall (ASQA) & 4 & 67.8 \\
& SELF-RAG & LLaMA2-7B & FactScore (ASQA) & 78 & 81.2 \\
& SELF-RAG & LLaMA2-13B & Precision (ASQA) & 2.3 & 70.3 \\
& SELF-RAG & LLaMA2-13B & Recall (ASQA) & 3.6 & 71.3 \\
& SELF-RAG & LLaMA2-13B & FactScore (ASQA) & 77.5 & 80.2 \\
& FiD-Light & T5-Base & FactScore (FEVER) & -- & 80.6 \\
& FiD-Light & T5-XL & FactScore (FEVER) & -- & 84.5 \\
& RAG w/ Rich Ans. Encoding & RAG & Recall (MSMARCO) & 25.3 & 27.5 \\
& RAG w/ Rich Ans. Encoding & RAG & Recall (KILT WoW) & 61.98 & 68.63 \\
& GenRT & RAG & Recall (NQ) & 59.4 & 60.78 \\
& GenRT & RAG & Recall (TriviaQA) & 68.22 & 70.01 \\
\midrule
\multirow{12}{*}{Hybrid RAG}
& CRAG & LLaMA2-7B & FactScore (Biography) & 59.2 & 74.1 \\
& Self-RAG & LLaMA2-7B & FactScore (Biography) & 59.2 & 81.2 \\
& Self-CRAG & LLaMA2-7B & FactScore (Biography) & 59.2 & 86.2 \\
& Flare Instruct & GPT-3.5 & Precision (2Wiki) & 48.6 & 49.1 \\
& Flare Instruct & GPT-3.5 & Recall (2Wiki) & 51.5 & 52.5 \\
& Flare Direct & GPT-3.5 & Precision (2Wiki) & 48.6 & 59.1 \\
& Flare Direct & GPT-3.5 & Recall (2Wiki) & 51.5 & 62.6 \\
& Stochastic RAG & FiD-Light (T5-Base) & FactScore (FEVER) & 80.6 & 81.3 \\
& Stochastic RAG & FiD-Light (T5-XL) & FactScore (FEVER) & 84.5 & 84.8 \\
& DRAGIN & LLaMA2-13B & Precision (HotPotQA) & 0.3711 & 0.4401 \\
& DRAGIN & LLaMA2-13B & Recall (HotPotQA) & 0.374 & 0.411 \\
& DRAGIN & VICUNA-13B & Precision (HotPotQA) & 0.3457 & 0.4226 \\
& DRAGIN & VICUNA-13B & Recall (HotPotQA) & 0.352 & 0.389 \\
\bottomrule
\end{tabular}
}
\end{table*}

\end{document}